\documentclass[12pt,a4paper,preprintnumbers,superscriptaddress,showkeys,]{revtex4} 

\usepackage[fleqn]{amsmath}
\usepackage{amssymb}
\usepackage{bm}
\usepackage{graphicx} 
\usepackage{subfigure}
\usepackage[pdfstartview=FitH,
            CJKbookmarks=true,
            bookmarksnumbered=true,
            bookmarksopen=true,
            colorlinks=true,
            pdfborder=001,
            linkcolor=blue,
            citecolor=blue,
            urlcolor=blue
            ]{hyperref}
\RequirePackage{color}
\usepackage{amsthm,amssymb,euscript}

\newtheorem{theorem}{Theorem}[section]

\newtheorem{lemma}[theorem]{Lemma}

\newcommand{\rf}{R}
\newcommand{\starB}{S}
 \newcommand{\sgn}{\textup{sgn}}

\begin{document}

\title{On the Decomposition  of Forces}

\author{Dong Eui Chang}
 \thanks{Email: dechang@uwaterloo.ca (old); dechang@kaist.ac.kr (current). Published in THEORETICAL \& APPLIED MECHANICS LETTERS {\bf 4}, 043001 (2014).}
\affiliation{
Department of Applied Mathematics,  University of Waterloo,
200 University Avenue West,
Waterloo, ON N2L 3G1, Canada
}%
%


\begin{abstract}
\noindent \textbf{Abstract} We show that any continuously differentiable  force is decomposed into the sum of a Rayleigh force and a gyroscopic force. We also extend this result to piecewise continuously differentiable forces. Our result improves the result on the decomposition of forces in a book by David Merkin  and further extends it to piecewise continuously differentiable forces. 
\end{abstract}


\maketitle

\section{Introduction}
Generally  forces depend on both position variables and velocity variables.  Among those forces, gyroscopic forces receive special attention since they do not contribute to   any change in the total energy of a mechanical system. Another special type of force is a Rayleigh force, which can be represented as the derivative of a function with respect to the velocity variables \cite{Ro60,Wh89}.  These two types of forces are important not only in mechanics  but also in control. For example, both gyroscopic forces  and dissipative Raleigh forces  are  employed in  feedback control design to stabilize mechanical systems \cite{Ch10:SIAM,NgChLa13}.   In this paper we show that any force, dependent upon both position and velocity,  is uniquely decomposed into the sum of a Rayleigh force and a gyroscopic force.  Constructive formulas and examples for the decomposition are provided.

\section{Main Results}

\subsection{Decomposition of Continuously Differentiable Forces}

Let $Q$ be a configuration manifold of dimension $n$. We denote by $TQ$ the tangent bundle, or velocity phase space, of $Q$ and by $T^*Q$ the cotangent bundle, or momentum phase space, of $Q$. We use $q = (q^1, \ldots, q^n)$ for coordinates on $Q$, and $(q,v) = (q^1,\ldots, q^n, v^1, \ldots, v^n)$ for coordinates on $TQ$.  Sometimes we simply write $v$  instead of $(q,v)$.  The cotangent bundle $T^*Q$ is locally spanned by  the coordinate one-forms $\{ \mathbf{d}q^1, \ldots, \mathbf{d}q^n\}$, where one can identify the set $\{ \mathbf{d}q^1, \ldots, \mathbf{d}q^n\}$ with the standard basis for $\mathbb R^n$ without loss of generality in this paper. The Einstein summation convention is employed such that repeated indices are implicitly summed over. 
Hence, we have the following identification:
\[
 a_i \mathbf{d}q^i  = a_1 \mathbf{d} q^1 + \cdots + a_n \mathbf{d}q^n = \begin{bmatrix} a_1\\ \vdots \\ a_n \end{bmatrix}.
\]

 A fiber-preserving map $F: TQ \rightarrow T^*Q$ is called a force map, or simply a force.  A force $F$ is called gyroscopic if  $\langle F(q,v), v\rangle = 0$ for all $(q,v) \in TQ$, where $\langle \, , \rangle$ is the canonical pairing between cotangent vectors in $T^*Q$ and tangent vectors in $TQ$.   A force $F$ is called a Rayleigh force if  there exists a function $\rf: TQ \rightarrow \mathbb R$ such that
 $F = \frac{\partial \rf}{\partial v^i} \mathbf{d}q^i$ for all $(q,v)\in TQ$,  where the function $\rf$ is called the Rayleigh function for the force $F$.

\begin{lemma}\label{rayleigh:lemma}
For any  continuously differentiable force $F: TQ \rightarrow T^*Q$, there exists a differentiable function $\rf: TQ \rightarrow \mathbb R$  such that
\begin{equation}\label{power:equal}
\langle F (q,v), v\rangle = \frac{\partial \rf}{\partial v^i}  v^i
\end{equation}
for all $(q,v) \in TQ$. Concretely, the function $\rf: TQ \rightarrow \mathbb R$ defined by
\begin{equation}\label{def:Pi}
\rf (q,v) = \int_0^1 \langle F(q,sv), v\rangle ds
\end{equation}
satisfies (\ref{power:equal}).  Moreover, if there is another function  $\tilde \rf : TQ \rightarrow \mathbb R$ such that
\begin{equation}\label{tilde:Pi}
\langle F (q,v), v\rangle = \frac{\partial \tilde \rf}{\partial v^i}  v^i
\end{equation}
for all $(q,v) \in TQ$, then there exists a function $f: Q \rightarrow \mathbb R$ such that
\[
\tilde \rf (q,v)=  \rf (q,v) +  f (q)
\]
for all $(q,v) \in TQ$.
\begin{proof}
Adapting the proof of the Proposition on  p.517 in \cite{MaHo93}, one can show that the function $\rf$ defined in (\ref{def:Pi}) is  differentiable on $TQ$. For any $\tau \in \mathbb R$
\begin{align*}
\rf (q,\tau v)  = \int_0^1 \langle F(q,s\tau v), \tau v\rangle ds = \int_0^\tau \langle F(q,sv), v\rangle ds.
\end{align*}
Hence,
\[
\frac{\partial \rf}{\partial v^i}v^i = \left . \frac{d }{d \tau}\right |_{\tau = 1} \rf(q,\tau v) = \langle F(q,v), v\rangle.
\]
Suppose that there is another differentiable function $\tilde \rf :TQ\rightarrow \mathbb R$ such that (\ref{tilde:Pi}) holds. 
Then,
\begin{align*}
\tilde \rf (q,v) &= \tilde \rf(q,0) + \int_0^1 \frac{d}{ds}\tilde \rf(q,sv) ds\\
&= \tilde \rf (q,0) + \int_0^1 \frac{\partial \tilde \rf}{\partial v^i} (q,sv) v^ids\\
&= \tilde \rf (q,0) + \int_0^1 \langle F(q,s v),  v\rangle ds\\
&=\tilde \rf (q,0) + \rf (q,v)
\end{align*}
where (\ref{tilde:Pi}) is used for the third equality. Therefore, the difference  between $\tilde \rf$ and $\rf$ is a function on $Q$.
\end{proof}
\end{lemma}

The following theorem states that any   continuously differentiable force can be expressed as the sum of a Rayleigh force and a  gyroscopic force. 
\begin{theorem}\label{theorem:decomposition:force}
For any continuously differentiable force $F: TQ \rightarrow T^*Q$, there exists a differentiable function $\rf: TQ \rightarrow \mathbb R$  and a  gyroscopic force $G: TQ \rightarrow T^*Q$ such that
\begin{equation}\label{FRG}
F = \frac{\partial \rf}{\partial v^i}\mathbf{d}q^i + G.
\end{equation}
 Moreover, the decomposition is unique.
 \begin{proof}
 Let $\rf $ be the function defined in (\ref{def:Pi}), and  let $G = F - \frac{\partial \rf}{\partial v^i}\mathbf{d}q^i$. By Lemma~\ref{rayleigh:lemma},  it is easy to show that $G$ is gyroscopic. Suppose that there exists another differentiable function $\tilde \rf$ on $TQ$ and another gyroscopic force $\tilde G$ such that
 \begin{equation}\label{FRG:tilde}
 F = \frac{\partial \tilde \rf}{\partial v^i}\mathbf{d}q^i + \tilde G.
 \end{equation}
 Since both $G$ and $\tilde G$ are gyroscopic, by (\ref{FRG}) and (\ref{FRG:tilde})
 \[
  \frac{\partial \rf}{\partial v^i}v^i = \langle F(q,v), v \rangle =  \frac{\partial \tilde \rf}{\partial v^i}v^i.
 \]
Hence, by Lemma~\ref{rayleigh:lemma} there exists a function $f: Q\rightarrow \mathbb R$ such that $\rf (q,v) = f(q) + \tilde \rf(q,v)$. Thus, $\frac{\partial \rf}{\partial v^i}\mathbf{d}q^i = \frac{\partial \tilde \rf}{\partial v^i}\mathbf{d}q^i$ and  $G = F - \frac{\partial \rf}{\partial v^i}\mathbf{d}q^i = F - \frac{\partial \tilde \rf}{\partial v^i}\mathbf{d}q^i = \tilde G$, which proves the uniqueness of  the decomposition.
 \end{proof}
\end{theorem}

Notice that  Theorem  \ref{theorem:decomposition:force} is an improvement of the exposition in Section 6.2 of \cite{Me97}. Our proof herein is more rigorous, and we discuss and prove the uniqueness of the decomposition as well.

Let us take an example. Consider a force 
\[
F(q,v) = \begin{bmatrix}
q^1 v^2 \\ v^1v^2
\end{bmatrix}
\]
on $\mathbb R^2$, where $q = (q^1,q^2)$ and $v=(v^1,v^2)$. By (\ref{def:Pi})
\begin{align*}
R(q,v) &= \int_0^1  (s q^1v^1v^2 + s^2 v^1 (v^2)^2 ) ds = \frac{1}{2} q^1v^1v^2 + \frac{1}{3} v^1 (v^2)^2,\\
\frac{\partial R}{\partial v} &= \begin{bmatrix} \frac{\partial R}{\partial v^1} \\ \frac{\partial R}{\partial v^2}\end{bmatrix} = \begin{bmatrix} \frac{1}{2}q^1v^2 + \frac{1}{3}(v^2)^2 \\ \frac{1}{2} q^1v^1 + \frac{2}{3}v^1v^2\end{bmatrix},\\
G &= F - \frac{\partial R}{\partial v} = \begin{bmatrix} \frac{1}{2}q^1 v^2 - \frac{1}{3}(v^2)^2 \\ \frac{1}{3}v^1v^2 - \frac{1}{2}q^1v^1\end{bmatrix}
\end{align*}
such that
\[
F(q,v) = \begin{bmatrix} \frac{1}{2}q^1v^2 + \frac{1}{3}(v^2)^2 \\ \frac{1}{2} q^1v^1 + \frac{2}{3}v^1v^2\end{bmatrix} + \begin{bmatrix} \frac{1}{2}q^1 v^2 - \frac{1}{3}(v^2)^2 \\ \frac{1}{3}v^1v^2 - \frac{1}{2}q^1v^1\end{bmatrix},
\]
where the first term on the right-hand side is the Raleigh  part and the second the gyroscopic part of $F$. 


Let us take another example. Consider a force  $F(q,v) = a(q)(v^1)^{i_1} \cdots (v^n)^{i_n} \mathbf{d}q^1$ on $\mathbb R^n$, where $i_k$'s are non-negative integers such that $\sum_{k=1}^n i_k \geq 0$. Then,  by (\ref{def:Pi}) 
\[
\rf(q,v) =  \frac{a(q)}{ (i_1 +1) + \cdots + i_n}(v^1)^{i_1+1} \cdots (v^n)^{i_n}.
\]
One can easily extend this computation to the case where  $F$ is a force each component of which is a polynomial in $v$, since  the left-hand side of (\ref{power:equal}) is linear in $F$ and the right-hand side of (\ref{power:equal}) is linear in $\rf$.

We now give an example to show the role of a gyroscopic force as a coupling force. Consider the following forced 2-dimensional harmonic oscillator:
\begin{align*}
\ddot q^1 + q^1 &= -\dot q^1 + \epsilon (\dot q^2)^2,\\
\ddot q^2 + q^2 &= -\epsilon \dot q^1 \dot q^2,
\end{align*}
where  $\epsilon$ is a constant parameter. One can easily show that the force vector on the right-hand side is decomposed as follows:
\[
F = \begin{bmatrix} -\dot q^1 \\ 0 \end{bmatrix} + \epsilon \begin{bmatrix}
 (\dot q^2)^2 \\ - \dot q^1 \dot q^2
\end{bmatrix},
\]
where the first vector on the right-hand side is a Rayleigh force and the second a gyroscopic force. The Rayleigh force is dissipative and the total energy function of the system is given by
\[
E = \frac{1}{2} ( (q^1)^2 + (q^2)^2 + (\dot q^1)^2 + (\dot q^2)^2).
\]
 If $\epsilon = 0$, then the $q^1$-dynamics and the $q^2$-dynamics get decoupled. As a result the $q^2$-dynamics becomes only Lypunov stable, not asymptotically stable, while the $q^1$-dynamics is exponentially stable.  However, if $\epsilon \neq 0$, then the two dynamics get coupled through the gyroscopic force, and one can apply LaSalle's invariance principle with the energy function $E$ as a Lyapunov function
to show that the origin is an asymptotically stable equilibrium point in the total dynamics. This shows that the gyroscopic force  creates a coupling between the two sub-dynamics and propagates the partially dissipative force throughout the entire dynamics.

\subsection{Decomposition of Piecewise Continuously Differentiable Forces} 

We now study  decomposition for a class of piecewise continuously differentiable forces. A good example is a Coulomb friction force such as  $F = -\sgn(v)$.  A submanifold $S$ of $TQ$ is called a star-shaped subbundle of $TQ$ if $ \starB = \bigcup_{q\in Q}\starB_q$ where  for each $q\in Q$ the set $\starB_q$ is a star-shaped open subset of  $T_qQ$, i.e., $\starB_q$ is an open subset of $T_qQ$ and $\lambda v_q \in \starB_q$ for all $v_q \in T_qQ$ and $0<\lambda\leq 1$.

\begin{theorem}\label{theorem:piecewise:C1:F}
Let $S$ be a star-shaped subbundle of $TQ$, and $F: TQ \rightarrow T^*Q$ be a force that is continuously differentiable on $S$. Let $x = (q,v) \in TQ$. Suppose that for each $x_0 \in \starB$ there exists a neighborhood $U$ of $x_0$ in $\starB$ such that the function $P: U \times (0,1] \rightarrow \mathbb R$ defined by $P((q,v),\tau ) = \langle F(q,\tau  v), v\rangle$  satisfies the following:
\begin{itemize}
\item [1.] $P$  extends to be continuous on $U \times [0,1]$,
\item [2.] $\frac{\partial}{\partial x}P$ extends to be continuous on $U \times [0,1]$.
\end{itemize}
Then, there exists a differentiable function $\rf :\starB \rightarrow \mathbb R$, and a  force $G: \starB \rightarrow T^*Q$ that is gyroscopic on $\starB$, i.e., $\langle G(q,v), v\rangle = 0$ for all $v\in \starB$, such that
\[
F = \frac{\partial \rf}{\partial v^i}\mathbf{d}q^i + G
\]
for all $(q,v) \in \starB$.
Moreover, the decomposition is unique.
\begin{proof}
By the hypothesis on $P(x,s)$, we may assume that $P$ and $\frac{\partial P}{\partial x}$ are continuous functions on $U\times [0,1]$. Then, adapting the Proposition in  p.517 of \cite{MaHo93}, one can show that
\[
\rf(q,v) = \lim_{\epsilon \rightarrow 0^+}\int_\epsilon^1 \langle F(q, \tau v), q\rangle d\tau
\]
is a well-defined differentiable function on $\starB$. The rest of the proof follows from the proof of Theorem~\ref{theorem:decomposition:force}.
\end{proof}
\end{theorem}



Here is an example. Let $Q = \mathbb R^2$ and $F = \cos(v^2) \sgn(v^1) \mathbf{d}q^1 + |v^2|\mathbf{d}q^2$. It is easy to see that $S =\{(q, v) \in \mathbb R^2 \times \mathbb R^2\mid v^1 \neq 0, v^2 \neq 0\}$ is a star-shaped subbundle of $TQ$. Then, for $\tau  \in (0,1]$ and $(q,v) \in S$
\begin{align*}
P((q,v),\tau):=\langle F(q,\tau v), v\rangle &= \cos(\tau v^2) \sgn(\tau v^1) v^1 + |\tau v^2|v^2\\
&= \cos(\tau v^2) |v^1| + \tau v^2 |v^2|,
\end{align*}
and
\begin{align*}
\frac{\partial P}{\partial q}((q,v),\tau) &=(0,0),\\
\frac{\partial P}{\partial v}((q,v),\tau ) &= ( \cos (\tau v^2) \sgn(v^1), -\tau \sin (\tau v^2)|v^1| + 2\tau |v^2|).
\end{align*}
It is easy to show that $P$, $\frac{\partial P}{\partial q}$ and $\frac{\partial P}{\partial v}$  extend to be continuous on $S \times [0,1]$, so the force decomposition is possible by Theorem \ref{theorem:piecewise:C1:F}. Indeed $F= \frac{\partial R}{\partial v^i}\mathbf{d}q^i + G$ where
\begin{align*}
\rf (q,v) &= \int_0^1(\cos( \tau v^2) |v^1| + \tau v^2 |v^2| ) d\tau  = \frac{\sin (v^2)}{v^2}|v^1| + \frac{1}{2}v^2 |v^2|,\\
\frac{\partial \rf}{\partial v^i}\mathbf{d}q^i &= \frac{\sin(v^2)}{v^2}\sgn(v^1) \mathbf{d}q^1 + \left (\frac{\cos(v^2) v^2 - \sin(v^2)}{(v^2)^2}|v^1| + |v^2| \right ) \mathbf{d}q^2,\\
G &= \frac{\cos(v^2)v^2 - \sin (v^2)}{v^2}\sgn(v^1) \mathbf{d}q^1 - \frac{\cos(v^2) v^2 - \sin(v^2)}{(v^2)^2}|v^1|\mathbf{d}q^2.
\end{align*}

\section{Conclusion}
We have shown that any continuously differentiable force can be uniquely expressed as the sum of a Rayleigh force and a gyroscopic force and extended this result to piecewise continuously differentiable forces. We have provided formulas for the decomposition and illustrated the formulas with examples.


\bibliographystyle{aipnum4-1}


\end{document}